\documentclass[aps,prl,twocolumn,groupedaddress,showpacs]{revtex4}

\usepackage{graphicx}

\begin{document}

\title{Long Phase Coherence Time and  Number Squeezing of two Bose-Einstein Condensates on an Atom Chip}

\author{G.-B. Jo, Y. Shin, S. Will, T. A.
Pasquini, M. Saba, W. Ketterle, and D. E. Pritchard}

\homepage[URL: ]{http://cua.mit.edu/ketterle_group/}

\affiliation{MIT-Harvard Center for Ultracold Atoms, Research
Laboratory of Electronics, Department of Physics, Massachusetts
Institute of Technology, Cambridge, MA 02139, USA}

\author{M. Vengalattore, M. Prentiss}

\affiliation{MIT-Harvard Center for Ultracold Atoms, Jefferson
Laboratory, Physics Department, Harvard University, Cambridge, MA
02138, USA}

\date{\today}

\begin{abstract}
We measure the relative phase of two Bose-Einstein condensates
confined in a radio frequency induced double well potential on an
atom chip. We observe phase coherence between the separated
condensates for times up to $\sim${200}~ms after splitting, a factor
of 10 longer than the phase diffusion time expected for a coherent
state for our experimental conditions. The enhanced coherence time
is attributed to number squeezing of the initial state by a factor
of 10. In addition, we demonstrate a rotationally sensitive (Sagnac)
geometry for a guided atom interferometer by propagating the split
condensates.
\end{abstract}


\maketitle Precision measurements in atomic physics are usually done
at low atomic densities to avoid collisional shifts and dephasing.
This applies to both atomic clocks and atom interferometers.  At
high density, the atomic interaction energy results in so-called
clock shifts~\cite{GCL93}, and leads to phase diffusion in
Bose-Einstein condensates
(BECs)~\cite{CDR97,LYQ96,WWG96,JWP97,LSC98,JWR98}. Most precision
measurements with neutral atoms are performed with free-falling
atoms in atomic beams~\cite{LHS97,GBK97} or in fountain
geometries~\cite{PCC99}. Major efforts are currently directed
towards atom interferometry using confined geometries, such as atom
traps or waveguides, often realized by using atom
chips~\cite{FKS02}. These geometries are promising in terms of
compactness and portability, and also offer the prospect of
extending interrogation times beyond the typical 0.5 s achievable in
the atomic fountains~\cite{PCC99}.

\begin{figure}
\begin{center}
\includegraphics{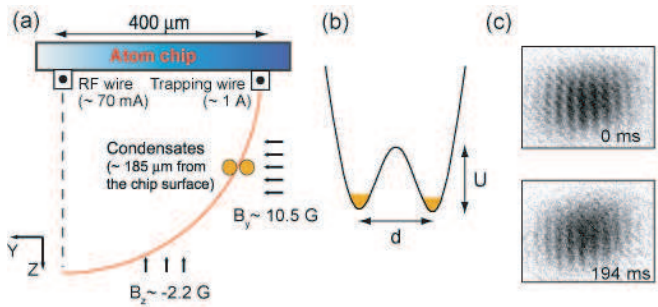}

\caption{Schematic of the atom chip interferometer. (a) Atoms were
confined radially by the combined magnetic potential of a
current-carrying wire and external bias field. Axial confinement in
$x$ direction was provided by a pair of endcap wires (not
shown)~\cite{SJP05}. By dressing the atoms with an oscillating radio
frequency (RF) field from a second wire, the single minimum in the
magnetic trapping potential was deformed into a
double-well~\cite{SHA05}. If the trapping position lies on the
circle containing the trapping wire and centered on the RF wire, the
splitting occurs in the horizontal plane. Condensates were placed
185~$\mu$m away from the chip surface. For the single well, the
radial (axial) trap frequency was $f_r=2.1$~kHz ($f_z=9$~Hz) and the
Larmor frequency at the trap center was $\approx 190$~kHz
($B_x\approx 0.27$~G). Splitting was performed over 75~ms by
linearly ramping the frequency of the RF field from 143~kHz to
225~kHz. Gravity points in the $+z$ direction. (b) Double-well
potential. The separation $d$ between the two wells and the barrier
height $U$ were controlled by adjusting the frequency or amplitude
of the RF field. (c) Matter wave interference. For various hold
times after splitting, absorption images of condensates released
from the double-well potential were taken after 10~ms
time-of-flight. The field of view is $260\times200~\mu$m.}
\end{center}
\end{figure}

However, given the deleterious effects of high atomic density, those
devices were thought to be able to operate only at low density and
therefore at small flux, seriously limiting the achievable
signal-to-noise ratio and sensitivity.  Here we show that we can
operate BEC interferometer at high density, with mean field energies
exceeding ${h}~\times$~{6}~kHz, where $h$ is Planck's constant.
Using a radio frequency (RF) induced beam splitter
~\cite{ZGT01,CKM04,SHA05}, we demonstrate that condensates can be
split reproducibly, so that even after 200~ms, or more than one
thousand cycles of the mean field evolution, the two condensates
still have a controlled phase. The observed coherence time of 200~ms
is ten times longer than the phase diffusion time for a coherent
state. Therefore, repulsive interactions during the beam splitting
process~\cite{MAC01} have created a non-classical squeezed state
with relative number fluctuations ten times smaller than for a
Poissonian distribution.

Our work is a major advance in the coherence time of confined atom
interferometers, which have operated at interrogation times below
$\sim${50}~ms~\cite{SSP04,SHA05,GDH06} due to technical limitations.
Our work also advances the preparation of number squeezed states to
much higher atom numbers.  Previous experiments in optical
lattices~\cite{OTF01,GME02} and in an optical trap~\cite{CSM05} were
limited to very small populations ($\sim1-1000$~atoms).  In
addition, the fact that the clouds could be prepared on an atom chip
with dc and RF electric currents, but without any laser beams, is
promising for future applications. Finally, operating the RF induced
beam splitter on propagating condensates, we realized an on-chip
Sagnac interferometer.

For two separated Bose-Einstein condensates, a state of well-defined
relative phase (phase coherent state), $|\phi\rangle$, is a
superposition state of many relative number states,
$|N_r=N_1-N_2\rangle$, where $N_1$ and $N_2$ are the occupation of
each well for $N = N_1+N_2$ atoms. Because of atom-atom interactions
in the condensates, the energy of number states, $E(N_1,N_2)$, have
quadratic dependence on the atom numbers $N_1$ and $N_2$ so that the
different relative number states have different phase evolution
rates~\cite{CDR97,JWP97}. A superposition state will therefore have
a spread of evolution rates, causing ``phase diffusion" or
``decoherence" of the relative phase with time. In contrast to
normal diffusion processes, the phase uncertainty, $\Delta\phi$,
increases here linearly. The phase diffusion rate, $R$, is
proportional to the derivative of the chemical potential of
condensates, $\mu(N_i)~(i=1,2)$, with respect to the atom number and
the standard deviation of the relative atom number, $\Delta N_r$:
$R=
(2\pi/h)(d\mu/dN_i)_{N_i=N/2}\Delta{N_r}$~\cite{CDR97,LYQ96,WWG96,JWP97,LSC98,JWR98}.
A number squeezed state with sub-Poissonian number fluctuations
($\Delta N_r = \sqrt{N}/s$), where $s>1$ is the squeezing factor,
will exhibit a reduced phase diffusion rate relative to a phase
coherent state with $\Delta N_r =\sqrt{N}$.

Bose-Einstein condensates of $\sim4\times10^{5}$ $^{23}$Na atoms in
the $|F=1, m_{F}=-1\rangle$ state were transferred into a magnetic
trap generated by the trapping wire on an atom chip and external
bias field~\cite{SJP05}. A double-well potential in the horizontal
plane was formed using adiabatic RF-induced splitting as described
in Fig.~1a~\cite{ZGT01,SHA05}. Typically, the separation of the two
wells was $d \sim 8.7~\mu$m, the height of the trap barrier was $U
\sim h \times 30$~kHz, and the chemical potential of the
condensates, measured from the trap bottom, was $\mu \sim h \times
6$~kHz (Fig. 1b). The life-time of the atoms at the splitting
position was $\sim${1.8}~s, significantly longer than in our
previously demonstrated two-wire splitting
method~\cite{HVB01,CKM04,SJP05}. RF-induced splitting has several
advantages over two-wire schemes: no loss channel (open port) during
splitting, less sensitivity to magnetic bias fields, and realization
of high trap-frequencies far from a surface~\cite{LSH06}. Atoms were
held in the double well for varying hold times, released by turning
off the trapping potential within {30}~$\mu$s at a known phase of
the RF field~\cite{foot:strongfield}, overlapped, and interfered in
time-of-flight (Fig.~1c). The relative phase of the two condensates
was measured as the spatial phase of the interference
pattern~\cite{SSP04,SJP05}.

\begin{figure}
\begin{center}
\includegraphics{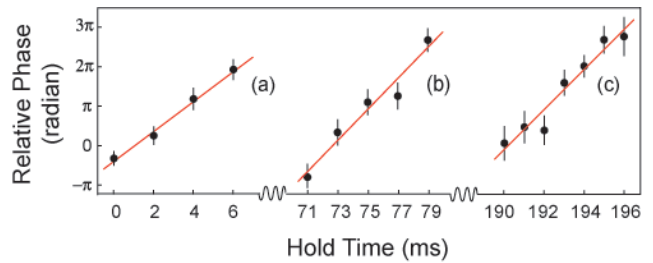}

\caption{Phase evolution of the relative phase during three
different time intervals. The evolution rate of the relative phase
are determined from the linear fit to be (a) 191~Hz, (b) 198~Hz, and
(c) 255~Hz. The data points represent the average of ten
measurements for (a) and (b), and fifteen for (c).}
\end{center}
\end{figure}

An atom interferometer requires two independent condensates without
any weak link which may lock the phase~\cite{SSP04, SJP05}.  To
demonstrate this independence, we monitored the relative phase over
short intervals right after and after up to {190}~ms delay time.
During each of these intervals, the phase evolved linearly with time
at $\sim2\pi\times200$~Hz, the signature of independent condensates
(Fig.~2).  The non-vanishing phase evolution rates are attributed to
small asymmetries in the two trapping potentials, which lead to
slightly different chemical potential after the splitting process.
The time variation of this rate is attributed to axial motion of the
two separated condensates. Note that the observed drift of the phase
evolution rate of $\sim$~60~Hz is only 1$\%$ of the condensates'
chemical potential. In principle, the phase drift could be zeroed by
a compensation field, but this has not been attempted.

To rule out the possibility that any weak link existed during the
200 ms time evolution and reset the relative phase, we demonstrated
that an applied phase shift could be read out 200 ms after its
appliance (Fig.~3). This proves that two independently evolving
condensates have preserved phase coherence up to {200}~ms, a factor
of 10 longer than the phase diffusion time,
$\tau_{c}=1/R\simeq${20}~ms, for our parameters.

\begin{figure}
\begin{center}
\includegraphics{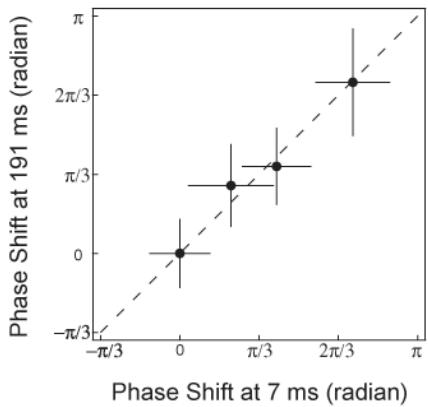}

\caption{Long phase coherence of two separated condensates. Applying
various phase shifts to the condensates at 2~ms after splitting, the
shifts of the relative phase were measured at 7~ms and 191~ms,
showing strong correlation. The dotted line denotes the ideal case
of perfect phase coherence. Phase shifts were applied by pulsing an
additional magnetic field in the $z$ direction for 50~$\mu$s with
variable amplitude.}
\end{center}
\end{figure}

For the quantitative study of phase fluctuations, the standard
deviation of the phase does not provide the best characterization
because the phase is measured modulo 2$\pi$. In the limit of a large
data set, a completely random distribution has a phase variance of
$\sim (3\pi/5)^2$. Already for smaller variances, the overlap of the
tails of the Gaussian distribution can cause ambiguities. As a more
appropriate measure of correlation, we represent each measurement of
the relative phase as a phasor with unit length and compare the
length of the sum of $N$ measured phasors with the expectation value
of $\sqrt{N}$ for $N$ random phasors. The larger the difference, the
smaller is the probability that the data set is compatible with a
random phase distribution. This probability of uncorrelated phases
is called randomness~\cite{FNS93}.
Uncorrelated data have an expected value of randomness near 0.5,
while strongly correlated data would have a small value, e.g. ten
data points drawn from a distribution with variance $(\pi/5)^2$ have
a probability of only $10^{-4}$ to be compatible with uncorrelated
phases.

\begin{figure}
\begin{center}
\includegraphics{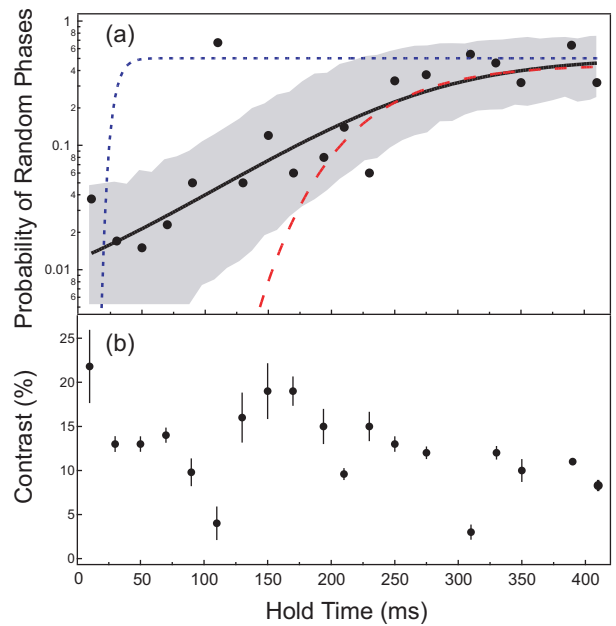}

\caption{Phase diffusion and number squeezing. (a) The randomness
probability of ten measurements of the relative phase is displayed
up to 400~ms after splitting. The blue dotted curve (red dashed
curve) shows a simulation for a phase-coherent state (number
squeezed state with $s=10$), which have negligible initial phase
uncertainty. The solid line includes an initial phase uncertainty of
0.28$\pi$ (see text). The shaded region represents the window where
ten data points from the sample with the given phase uncertainty
would fall with 50$\%$ probability. (b) Contrast of the interference
pattern. Since the endcap wires generate a field gradient
$\frac{\partial B_{z}}{\partial x}$ as well as a field curvature
$\frac{{\partial}^{2} B_{x}}{\partial x^{2}}$ at the position of the
condensates, the two wells are not parallel to the trapping wire and
consequently have slightly different axial trapping potential. This
difference induces relative axial motion of the two condensates,
which periodically reduces the contrast.}
\end{center}
\end{figure}

To study phase diffusion in our system, we analyze the distribution
of ten measurements of the relative phase at various times after
splitting, as shown in Fig.~4.  The data show a well-defined phase
(with probability $>$~90$\%$) for times shorter than $\sim200$~ms.
In contrast, the simulation for a coherent state in our experimental
conditions, shown as a blue dotted line, predicts the same scatter
of phase measurements already after $\sim${20}~ms. Fitting a phase
diffusion model to the data points with randomness probability
$>$~0.1, gives a phase diffusion time of 200~ms. The solid line is a
fit which includes the initial variance $\Delta \phi_{0}^2$:
\begin{equation}
\Delta\phi(t)^{2}=\Delta\phi_{0}^{2}+(Rt)^{2}
\end{equation}
The variance of the initial state, $\Delta \phi_0 ^2 = (0.28\pi)^2$
is dominated by technical noise including fitting errors and
non-ideal trap switch off. The contribution due to initial number
fluctuations,
 $\Delta\phi_{0}^2\simeq(s/\Delta
N_r)^2$, is $\sim(7.1\times10^{-4}\pi)^2$  for a coherent state
($s$=1), and remains small unless the squeezing leads to number
fluctuations on the order of a single atom, $s\sim\sqrt{N}$. The
fitted value for the phase diffusion rate of $R=5~s^{-1}$ includes
technical shot-to-shot variations in the relative atom number of two
condensates after splitting and thermal fluctuation. Therefore, the
inferred squeezing factor $s$=10 represents a lower bound.  It
implies that our relative atom number fluctuations were smaller than
$\pm0.03\%$ corresponding to $\pm$50 atoms.

Locally the interference pattern of two pure condensates should
always have 100$\%$ contrast, where contrast is defined as the
density amplitude of the interference fringes over the mean density.
Since in our experiment the contrast is derived from interference
pattern integrated along the line-of-sight,
 it decreased gradually with time and exhibited fluctuations most likely due to
asymmetric axial motion~(Fig.~4b). Except for small regions near 110
and 300~ms hold time, the contrast was above 10$\%$, sufficient for
accurate determination of the phase. The small windows with poor
contrast have a large probability for random phases.

The observed long phase coherence time implies that the initial
state is number squeezed. The probable origin of number squeezing
during the splitting is repulsive atom-atom
interactions~\cite{MAC01}. The interactions make it energetically
favorable for the two condensates to split with equal numbers in a
symmetric double-well potential, whereas number fluctuations , such
as in a coherent state, cost energy. Describing splitting dynamics
by the Josephson Hamiltonian shows how the interplay of tunneling
and interactions leads to an increase of squeezing as the barrier is
increased~\cite{MAC01,BLD06}. Assuming that the squeezing can no
longer increase when the Josephson frequency becomes slower than the
inverse splitting time, we estimate a squeezing factor of $\sim$13
for our experimental conditions~\cite{BLD06}. For our elongated
condensates, phase fluctuations are present for temperature above
$\sim${100}~nK which is $\sim$1/10 of the BEC transition
temperature~\cite{PSW01}. Since we cannot measure the temperature of
an almost pure condensate in time of flight, it is not clear whether
the interfering condensates had correlated phase fluctuations or
not.

\begin{figure}
\begin{center}
\includegraphics{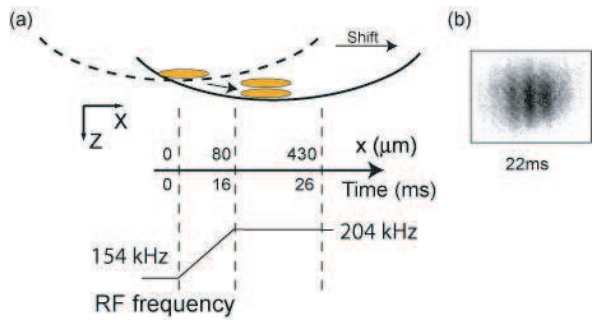}

\caption{ Confined atom interferometry with enclosed area. (a) A
single trapped condensate at rest was accelerated by shifting the
trap center by $\sim 430~\mu$m in the axial direction ($t=0$~ms).
The frequency of the RF field was linearly ramped from 154~kHz to
204~kHz during 16~ms after launch, splitting the condensates and
separating them by $5-6~\mu$m. (b) Phase measurements were done for
up to 26~ms after the launch. The probability for random phases was
determined for data sets of ten measurements. Until 16~ms, this
probability was extremely low (less than $10^{-12}$) and the
relative phase was constant, implying that the two condensates were
still connected. For $t>16$~ms, the relative phase evolved (similar
to Fig. 2), and the probability for a random phase distribution was
smaller than 10$\%$. This demonstrates that phase coherence was
preserved after full splitting.  The figure shows the interference
pattern for $t=22$~ms. The field of view is
260$\times${200}~$\mu$m.}
\end{center}
\end{figure}

The work presented so far, and also all previous work on
interferometry of confined or guided atoms, featured geometries
without any enclosed area between the two paths of the
interferometer.  An enclosed area is necessary for rotational
sensitivity ~\cite{SSC13,LHS97,GBK97} and  requires moving atoms. As
described in Fig.~5, we were able to extend the coherent beam
splitter to condensates moving on an atom chip.  The observed
coherence time (10 ms) and  propagation distance after splitting
($\sim350~\mu$m) were only limited by the chip geometry. This
corresponds to an enclosed area of $\sim${1500}~$\mu m^{2}$, and a
response factor $\frac{4\pi m A}{h} \sim 7.9\times10^{-5}
rad/\Omega_{e}$ for rotation sensing, where $m$ is the probe
particle mass and $\Omega_{e}$ the earth rotation rate~\cite{BPA97}.

In conclusion, the present work demonstrates a long phase coherence
time of $\sim$200~ms between two spatially separated condensates on
an atom chip, rivaling the interrogation times in fountain-type
interferometers~\cite{PCC99}. Number squeezing by a factor $\geq10$
occurs during the preparation of the split state, providing a
well-defined phase beyond the phase diffusion limit for a coherent
state. Thus, interaction-induced squeezing reduces the phase
diffusion caused by the same interactions~\cite{MAC01}. These
results show that it is both possible and promising to use
condensates at high density for interferometry on an atom chip.

This work was funded by DARPA, NSF, ONR, and NASA. G.-B. Jo and S.
Will acknowledge additional support from the Samsung Scholarship and
the Studienstiftung des deutschen Volkes, respectively. We thank
C.Christensen for experimental assistance.


\begin{thebibliography}{27}
\bibitem{foot:strongfield}
We observed strong correlation between the population distribution
of the spin components and the phase of the RF field at the moment
of release, probably since the strength of the RF field
($\sim$0.35G) was comparable to the local static field
($\sim$0.27G)~\cite{ATS55}. The trapping potential was switched off
at a value of the RF phase chosen to release atoms predominantly in
the $|F=1, m_{F}=-1\rangle$ state.
\end{thebibliography}

\end{document}